\documentclass{emulateapj}

\slugcomment{Draft Version Aug. 10, 2006: Accepted for Publication in ApJ}

\usepackage{times}
\usepackage{bm}% bold math
\usepackage{amsmath,amssymb}

\newcommand{\bolB}{{\bm  B}}
\newcommand{\bolJ}{{\bm  J}}
\newcommand{\bolV}{{\bm  V}}

\shorttitle{Structure of Magnetic Tower Jets}
\shortauthors{Nakamura, Li, \& Li}

\begin{document}

\title{Structure of Magnetic Tower Jets in Stratified Atmospheres}

\author{Masanori Nakamura\altaffilmark{1}, Hui Li\altaffilmark{1}, 
and Shengtai Li\altaffilmark{2}}
  
\altaffiltext{1}{Theoretical Astrophysics, MS B227, Los Alamos
  National Laboratory, NM 87545; nakamura@lanl.gov; hli@lanl.gov}
\altaffiltext{2}{Mathematical Modeling and Analysis, MS B284, Los
  Alamos National Laboratory, NM 87545; sli@lanl.gov}

\begin{abstract}

Based  on  a  new  approach  on modeling  the  magnetically  dominated
outflows  from AGNs  (Li et  al. 2006),  we study  the  propagation of
magnetic tower jets in gravitationally stratified atmospheres (such as
a galaxy  cluster environment)  in large scales  ($>$ tens of  kpc) by
performing  three-dimensional  magnetohydrodynamic (MHD)  simulations.
We present  the detailed  analysis of the  MHD waves,  the cylindrical
radial force balance,  and the collimation of magnetic  tower jets. As
magnetic energy is injected into  a small central volume over a finite
amount of time, the magnetic fields expand down the background density
gradient, forming a collimated jet and an expanded ``lobe'' due to the
gradually decreasing background density and pressure. Both the jet and
lobes  are magnetically  dominated.   In addition,  the injection  and
expansion produce  a hydrodynamic shock  wave that is moving  ahead of
and enclosing the magnetic tower  jet. This shock can eventually break
the hydrostatic equilibrium  in the ambient medium and  cause a global
gravitational  contraction.    This  contraction  produces   a  strong
compression  at the  head of  the magnetic  tower front  and  helps to
collimate radially to produce a slender-shaped jet.  At the outer edge
of the jet, the magnetic pressure is balanced by the background
(modified) gas pressure, without any significant contribution from the
hoop stress.   On the other hand,  along the central axis  of the jet,
hoop stress is  the dominant force in shaping  the central collimation
of the poloidal  current.  The system, which possesses  a highly wound
helical  magnetic  configuration,  never  quite reaches  a  force-free
equilibrium  state though the  evolution becomes  much slower  at late
stages.    The  simulations   were  performed   without   any  initial
perturbations  so the  overall  structures of  the  jet remain  mostly
axisymmetric.

\end{abstract}


%\sloppy

\section{INTRODUCTION}

A number of astronomical systems have been discovered to eject tightly
collimated and hyper-sonic plasma  beams and large amounts of magnetic
fields  into the interstellar,  intracluster and  intergalactic medium
from the  central objects  during their initial/final  (often violent)
stages.  Magnetohydrodynamic  (MHD) mechanisms are  frequently invoked
to  model the  launching, acceleration  and collimation  of  jets from
Young Stellar  Objects (YSOs), X-ray binaries  (XRBs), Active Galactic
Nuclei  (AGNs),  Microquasars,  and  Quasars (QSOs)  \citep[see,  {\it
e.g.},][and references therein]{M01}.

Theory of  magnetically driven outflows in  the electromagnetic regime
has been proposed by  \citet[]{B76} and \citet[]{L76} and subsequently
applied  to  rotating black  holes  \citep[]{BZ77}  and to  magnetized
accretion  disks   \citep[]{BP82}.   By  definition,   these  outflows
initially are dominated by electromagnetic forces close to the central
engine.   In  these  and  subsequent  models  of  magnetically  driven
outflows (jets/winds), the plasma velocity passes successively through
the  hydrodynamic  (HD)   sonic,  slow-magnetosonic,  Alfv\'enic,  and
fast-magnetosonic critical surfaces.

The  first  attempt  to  investigate  the  nonlinear  (time-dependent)
behavior  of magnetically  driven  outflows from  accretion disks  was
performed by \citet[]{US85}.  The  differential rotation in the system
(central star/black  hole and the  accretion disk) creates  a magnetic
coil  that simultaneously  expels and  pinches some  of  the infalling
material.  The buildup of  the azimuthal (toroidal) field component in
the  accretion disk  is released  along  the poloidal  field lines  as
large-amplitude   torsional   Alfv\'en   waves  (``sweeping   magnetic
twist'').   After  their  pioneering   work,  a  number  of  numerical
simulations  to study  the MHD  jets have  been done  \citep[see, {\it
e.g.},][and  references   therein]{F98}.   An  underlying  large-scale
poloidal field  for producing the  magnetically driven jets  is almost
universally  assumed in  many theoretical/numerical  models.  However,
the origin and  existence of such a galactic  magnetic field are still
poorly understood.

In   contrast   with  the   large-scale   field  models,   Lynden-Bell
\citep[]{LB94,  L96, L03,  L06} examined  the expansion  of  the local
force-free magnetic loops anchored to  the star and the accretion disk
by using  the semi-analytic approach.  Twisted magnetic  fluxes due to
the disk rotation make the magnetic  loops unstable and splay out at a
semi-angle 60\degr  \ from  the rotational axis  of the  disk.  Global
magnetostatic  solutions  of  magnetic  towers with  external  thermal
pressure were also computed by \citet[]{Li01} using the Grad-Shafranov
equation  in axisymmetry \citep[see  also,][]{L02, LR03,  UM06}.  Full
MHD numerical  simulations of magnetic  towers have been  performed in
two-dimension  (axisymmetric) \citep[]{R98, T99,  U00, K02,  K04a} and
three-dimension \citep[]{K04b}.  Magnetic  towers are also observed in
the laboratory experiments \citep[]{HB02, L05}.

This  paper describes  the  nonlinear dynamics  of propagating  magnetic
tower jets in large scales  ($>$ tens of kpc) based on three-dimensional
MHD  simulations.  We  follow closely  the approach  described in  Li et
al. (2006; hereafter Paper I). Different from Paper I, which studied the
dynamics of magnetic field evolution  in a uniform background medium, we
present  results  on  the  injection  and the  subsequent  expansion  of
magnetic  fields  in  a  stably  stratified background  medium  that  is
described  by  an  iso-thermal  King  model  \citep[]{K62}.   Since  the
simulated  magnetic structures  traverse  several scale  heights of  the
background medium, we  regard that our simulations can  be compared with
the radio sources inside the galaxy cluster core regions. Due to limited
numerical  dynamic range, however,  the injection  region (see  Paper I)
assumed in this  paper will be large  (a few kpc).  Our goal  here is to
provide the  detailed analysis of the  magnetic tower jets,  in terms of
its  MHD wave  structures,  its cylindrical  radial  force balance,  and
collimation.  The paper is organized as follows: In \S 2, we outline the
model  and numerical  methods.   In  \S 3,  we  describe the  simulation
results.  Discussions and conclusions are given in \S 4 and \S 5.

\section{MODEL ASSUMPTIONS AND NUMERICAL METHODS}

The basic model assumptions and numerical treatments we adopt here are
essentially the same as those  in Paper I.  Magnetic fluxes and energy
are  injected  into a  characteristic  central  volume  over a  finite
duration.   The injected  fluxes are  not force-free  so  that Lorentz
forces cause  them to expand, interacting with  the background medium.
For the  sake of completeness, we  show the basic  equations and other
essential numerical setup here again, and refer readers to Paper I for
more details.

\subsection{Basic Equations}

We solve  the nonlinear system  of time-dependent ideal  MHD equations
numerically in a 3-D Cartesian coordinate system ($x,\,y,\,z$):
\begin{eqnarray}
\label{eq:mass}
\frac{\partial\rho}{\partial t} + \nabla\cdot(\rho\mbox{\bf V}) &=& 
\dot{\rho}_{\rm inj}
\\
\label{eq:momentum}
\frac{\partial (\rho {\bf V})}{\partial t} + 
\nabla\cdot \left( \rho {\bf V V} + p + B^2/2 - 
{\bf B B}\right)  &=& - \rho \nabla\psi
\\
\label{eq:energy}
\frac{\partial E}{\partial t} + \nabla\cdot\left[\left(E
+p + B^2/2 \right)
{\bf v}-{\bf B}({\bf v}\cdot{\bf B})\right] &=& -\rho \bolV \cdot
\nabla \psi \nonumber \\ 
&+& \dot{E}_{\rm inj}\\
\label{eq:induction}
\frac{\partial {\bf B}}{\partial t} - \nabla\times( {\bf V}
\times {\bf B}) &=& {\dot {\bf B}}_{\rm inj},
\end{eqnarray}
%\beqn
%\label{eq:mass}
%&&\frac{\partial \rho}{\partial t} + \nabla \cdot (\rho \bolV) =
%\dot{\rho}_{\rm inj},\\
%\label{eq:momentum}
%&&\frac{\partial (\rho \bolV)}{\partial t} + \nabla \cdot \left[ 
%\rho \bolV \otimes \bolV - \frac{\bolB \otimes \bolB}{4 \pi} 
%+ \left(p+\frac{B^2}{8 \pi}\right) \bolI \right] = - \rho \nabla \psi, \\
%\label{eq:induction}
%&&\frac{\partial \bolB}{\partial t} + \nabla \times
%({\bolV}\times{\bolB})= \dot{\bolB}_{\rm inj},\\
%\label{eq:energy}
%&&\frac{\partial E}{\partial t} + \nabla \cdot \left[\left(
%\frac{\gamma p}{\gamma-1}+\frac{1}{2} \rho V^2 \right) \bolV -
%\frac{(\bolV \times \bolB) \times \bolB}{4 \pi}\right]
%=-\rho \bolV \cdot \nabla \psi + \dot{E}_{\rm inj}.
%\eeqn
Here $\rho$, $p$,  $\bolV$, $\bolB$, and $E$ denote  the mass density,
hydrodynamic (gas) pressure, fluid velocity, magnetic field, and total
energy,   respectively.   The   total   energy  $E$   is  defined   as
$E=p/(\gamma-1)+\rho V^2/2+B^2/2$, where $\gamma$  is the ratio of the
specific heats (a  value of $5/3$ is used).   The Newtonian gravity is
$-\nabla \psi$.   Quantities $\dot{\rho}_{\rm inj}$, $\dot{\bolB}_{\rm
inj}$,   and  $  \dot{E}_{\rm   inj}$  represent   the  time-dependent
injections of  mass, magnetic flux, and energy,  whose expressions are
given in Paper I.

We normalize  physical quantities with the unit  length scale $R_{0}$,
density  $\rho_{0}$,  velocity  $V_{0}$   in  the  system,  and  other
quantities  derived  from  their  combinations, {\it  e.g.},  time  as
$R_0/V_0$,  etc.  These  normalizing factors  are summarized  in Table
\ref{tbl:unit}.   Hereafter,  we  will  use the  normalized  variables
throughout the paper. Note that a  factor of $4 \pi$ has been absorbed
into  the  scaling for  both  the  magnetic  field ${\bolB}$  and  the
corresponding  current  density  ${\bolJ}$.   To  put  our  normalized
physical quantities in an astrophysical context, we use the parameters
derived  from the  X-ray observations  of  the Perseus  cluster as  an
example   \citep{C03}.   These   values  are   also  given   in  Table
\ref{tbl:unit}.

The system of  dimensionless equations is integrated in  time by using
an upwind  scheme \citep[]{LL03}.  Computations were  performed on the
parallel Linux clusters at LANL.

\subsection{Initial and Boundary Conditions}

One key difference from Paper I is that we now introduce a non-uniform
background  medium.   An  iso-thermal  model  \citep[]{K62}  has  been
adopted to model a gravitationally stratified ambient medium.  This is
applicable, for example, to modeling  the magnetic towers from AGNs in
galaxy clusters.

The  initial distributions of  the background  density $\rho$  and gas
pressure $p$ are assumed to be
\begin{equation}
\label{eq:king}
\rho = p = \left[1 + \left(\frac{R}{R_{\rm c}}\right)^2\right]^{-\kappa},
\end{equation}
where $R=(x^2+y^2+z^2)^{1/2}$ is the  spherical radius and $R_{\rm c}$
the  cluster   core  radius.   (In  the  following   discussion,  both
``transverse'' and ``radial'' have  the same meaning, referring to the
cylindrical  radial direction.)  The parameter  $\kappa$  controls the
gradient  of the  ambient  medium.  Furthermore,  we  assume that  the
ambient gas  is initially in  a hydrostatic equilibrium under  a fixed
(in  time  and space)  yet  distributed  gravitational field  $-\nabla
\psi(R)$ (such  as that generated  by a dark matter  potential).  From
the initial equilibrium, we get
\begin{equation} 
-\nabla \psi = \frac{\nabla
p|_{t=0}}{\rho|_{t=0}} = -\frac{2 \kappa R}{R_{\rm
c}^2}\left[1+\left(\frac{R}{R_{\rm c}^2} \right)^2\right]^{-1}.  
\end{equation}
%so the gravitational potential (to within an additive constant) is
%\begin{equation}
%\psi = - \ln p|_{t=0}~~.
%\end{equation}
In the present paper, we choose $R_{\rm c}=4.0$ and $\kappa=1.0$.  The
initial   sound   speed   in   the   system   is   constant,   $C_{\rm
s}|_{t=0}={\gamma}^{1/2}  \approx 1.29$, throughout  the computational
domain.   An important  time scale  is the  sound crossing  time $\tau
(\equiv  R/C_{\rm s})  \approx 0.78$,  normalized with  $\tau_{\rm s0}
(\equiv R_{0}  / C_{\rm  s0}) \approx 10.0$  Myrs.  Therefore,  a unit
time scale $t=1$ corresponds to $12.8$ Myrs.

The total computational domain is taken to be $|x| \leq 16$, $|y| \leq
16$, and $|z|  \leq 16$ corresponding to a $(80\  {\rm kpc})^3$ box in
the  actual  length  scales.   The  numbers  of  grid  points  in  the
simulations  reported  here  are  $N_{x}\times  N_{y}\times  N_{z}=240
\times 240 \times  240$, where the grid points  are assigned uniformly
in  the $x$,  $y$, and  $z$ directions.   A cell  $\Delta  x\ (=\Delta
y=\Delta z \sim 0.13)$ corresponds  to $\sim 0.65\ {\rm kpc}$.  We use
the outflow boundary conditions at all outer boundaries. Note that for
most of  the simulation duration,  the waves and magnetic  fields stay
within the simulation box,  and all magnetic fields are self-sustained
by their internal currents.

\subsection{Injections of Magnetic Flux, Mass, and Energy}

The injections of magnetic flux,  mass and its associated energies are
the  same  as those  described  in Paper  I.   The  ratio between  the
toroidal to poloidal fluxes of the injected fields is characterized by
a parameter  $\alpha = 15$, which  corresponds to $\sim  6$ times more
toroidal flux  than poloidal flux.  The magnetic  field injection rate
is described by $\gamma_b$ and is  set to be $\gamma_b=3$. The mass is
injected at a rate of $\gamma_\rho  =0.1$ over a central volume with a
characteristic  radius $r_\rho =  0.5$. Magnetic  fields and  mass are
continuously  injected  for  $t_{\rm  inj}  = 3.1$,  after  which  the
injection is  turned off.  These parameters correspond  to an magnetic
energy injection rate of $\sim  10^{43}$ ergs/s, a mass injection rate
of $\sim 0.046 M_\odot$/yr, and an injection time $\sim 40$ Myrs.

In summary, we have set  up an initial stratified cluster medium which
is in a  hydrostatic equilibrium.  The magnetic flux  and the mass are
steadily injected in a central small volume with a radius of 1.  Since
these magnetic fields  are not in a force-free  equilibrium, they will
evolve,  forming a  magnetic tower  and interacting  with  the ambient
medium.

\section{RESULTS}

In this section we examine  the nonlinear evolution and the properties
of magnetic tower jets in the gravitationally stratified atmosphere.

\subsection{Overview of Formation and Propagation of a
 Magnetic Tower Jet}

\begin{figure*}
\centerline{{\includegraphics[scale=1.1, angle=0]{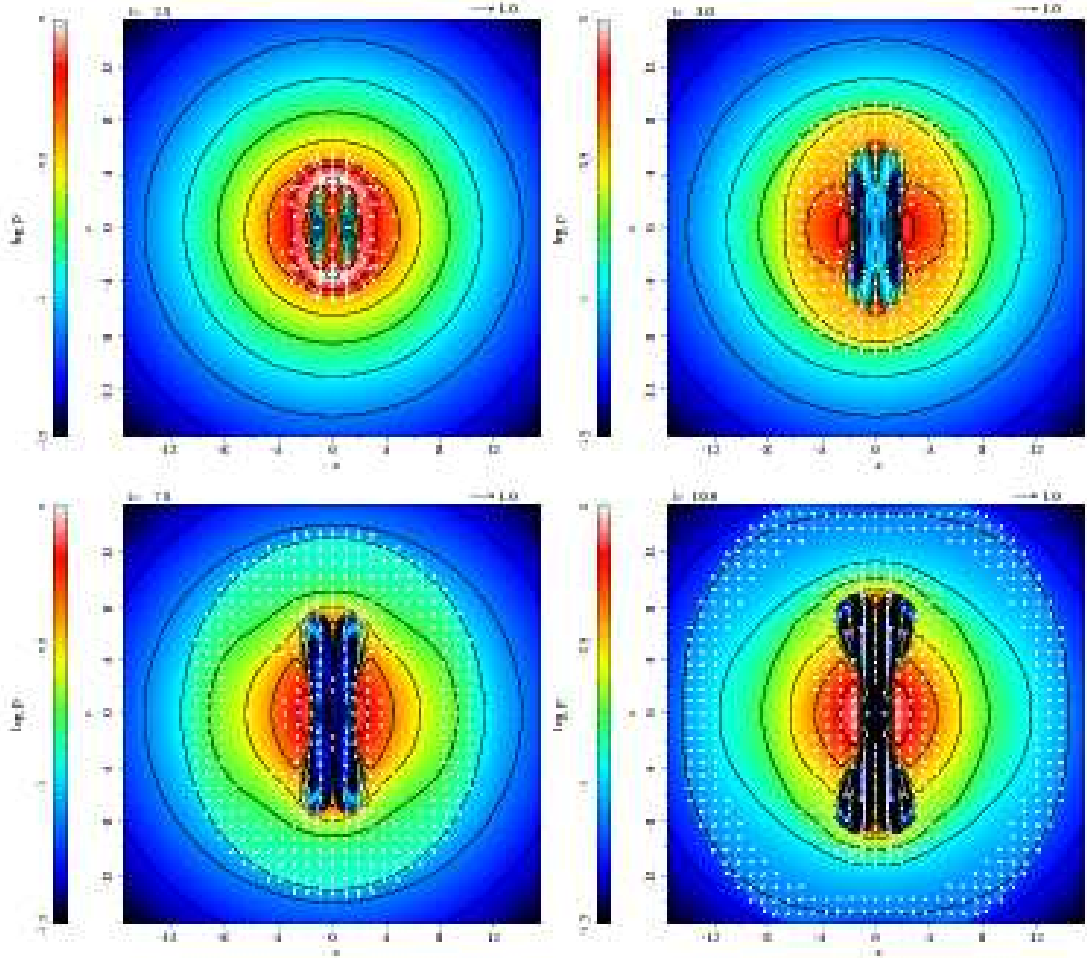}}}
\caption{\label{fig:de_evo} The density evolution caused by a magnetic
tower jet  propagating through  a stratified background.  Color images
and contours of the density distribution (logarithmic scale) are shown
along  with  the  poloidal   velocity  field  (arrows)  in  the  $x-z$
plane.  Density  cavities  are   formed  due  to  the  magnetic  field
expansion.  Times are  given at  the upper  left in  each of  the four
panels.  The  length of  the arrow  at the upper  right in  each panel
shows the unit velocity.  }
\end{figure*}

Before considering our numerical  results in detail, it is instructive
to give a brief overview of the time development of the magnetic tower
jet  system.  We  achieve  this by  presenting  the selected  physical
quantities   using  two-dimensional  $x-z$   slices  at   $y=0$.   The
distributions of  density at various  times $(t=2.5,\,5.0,\,7.5,\,{\rm
and}\,10.0)$ are  shown in Fig. \ref{fig:de_evo}.  At  the final stage
($t=10.0$), we  see the formation of a  quasi-axisymmetric (around the
jet axis)  magnetic tower jet with  low density cavities  (a factor of
$\sim 30$ smaller than the  peak density).  Inside these cavities, the
Alfv\'en speed is large $V_{\rm  A} \gtrsim 5.0$, while plasma $\beta$
is small  ($\beta = 2p/B^2  \lesssim 0.1$).  However,  the temperature
$T$ ($\propto  C_{\rm s}^2$)  becomes large $T  \gtrsim 2.5  $ ($\sim$
twice the initial constant value); that is, the hotter gas is confined
in  the  low-$\beta$ magnetic  tower.   The  jet  possesses a  slender
hourglass-shaped structure with a  radially confined ``body'' for $|z|
\leq 3$, which is likely due to the background pressure profile having
a  core  radius  $R_c =  4$.  As  the  magnetic  tower moves  into  an
increasingly  lower pressure  background, the  expanded  ``lobes'' are
formed.  A  quasi-spherical hydrodynamic  (HD) shock wave  front moves
ahead of the magnetic tower.

\begin{figure}
\begin{center}
\includegraphics[scale=0.7, angle=-90]{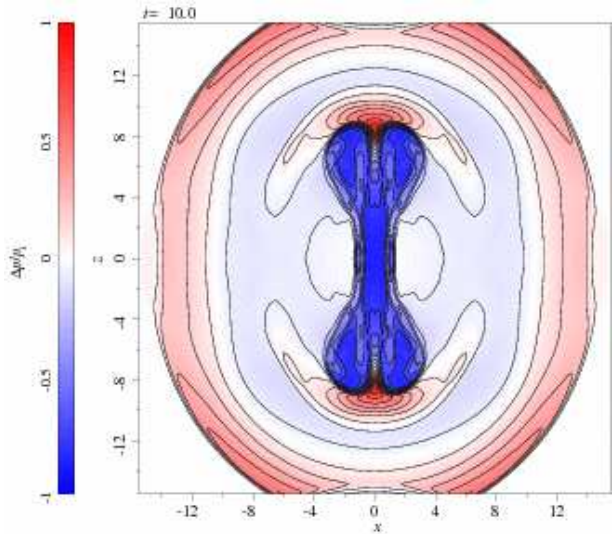}
\caption{\label{fig:final_prdiff}  The  change  of  the  gas  pressure
$\Delta  p/p_{\rm i}~(\Delta p=p-p_{\rm  i})$ ($p_{\rm  i}$ represents
$p|_{t=0}$) at $t=10.0$ in the $x-z$ plane.  }
\end{center}
\end{figure}

A snapshot of the gas pressure change ratio $\Delta p/p_{\rm i} =
p/p_{\rm i} -1$ (where $p_{\rm i}$ represents $p|_{t=0}$) at $t=10.0$
is shown in Fig. \ref{fig:final_prdiff}.  Positive $\Delta p$ can be
seen at both the post-shock region of the propagating HD shock wave
and just ahead of the magnetic tower ($|z| \sim 8-10$).  The
distribution of $\Delta p$ forms a ``{\sf U}''-shaped bow-shock-like
structure around the head of the magnetic tower.  This structure,
however,  does not  appear until  $t  \sim 7.5$.   This is  apparently 
caused  by the  local compression  between the  head of  the magnetic
tower and the reverse MHD slow-mode wave (see discussions in the next
section).  The gas pressure inside the magnetic tower becomes small
($|\Delta p/p_{\rm i}| \lesssim 0.5$) due to the magnetic flux
expansion.  Note that the light-blue region between the HD shock and
the magnetic tower shows a small pressure decrease ($\Delta p/p_{\rm
i} \approx - 0.1$).  In later sections, we will discuss the origin of
this depression and what role it plays in the dynamics of magnetic
tower jets.

\begin{figure}
\begin{center}
\includegraphics[scale=1.2, angle=0]{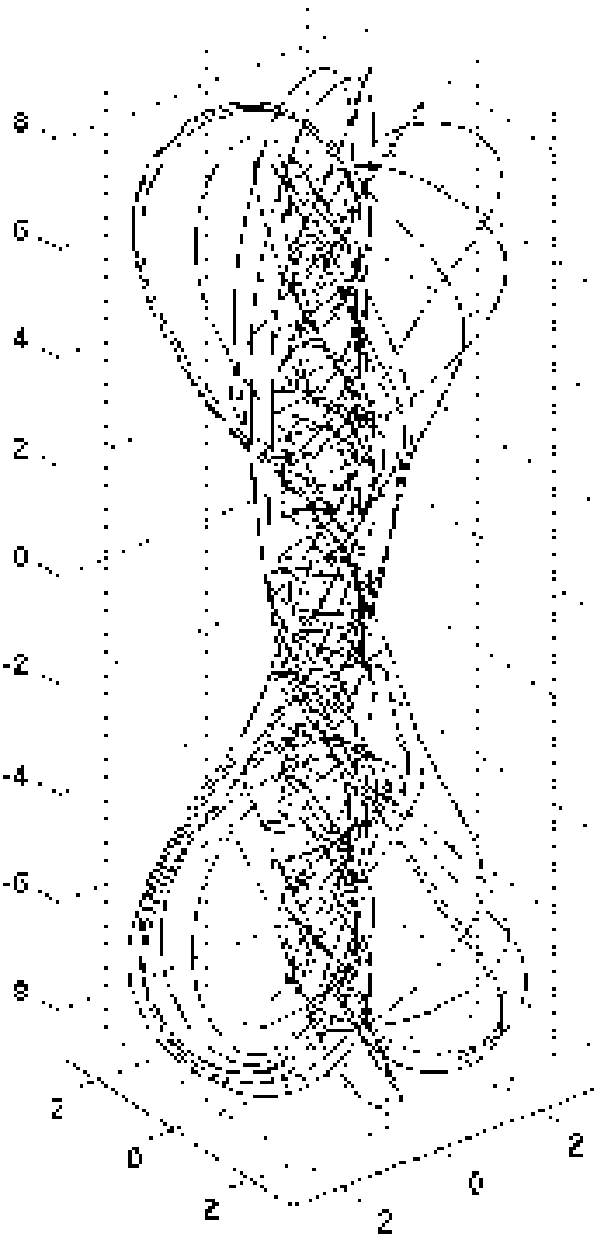}
\caption{\label{fig:3Dlines}  Three-dimensional configuration  view of
the selected magnetic lines at $t=10.0$.  }
\end{center}
\end{figure}

\begin{figure}
\begin{center}
\includegraphics[scale=0.7, angle=-90]{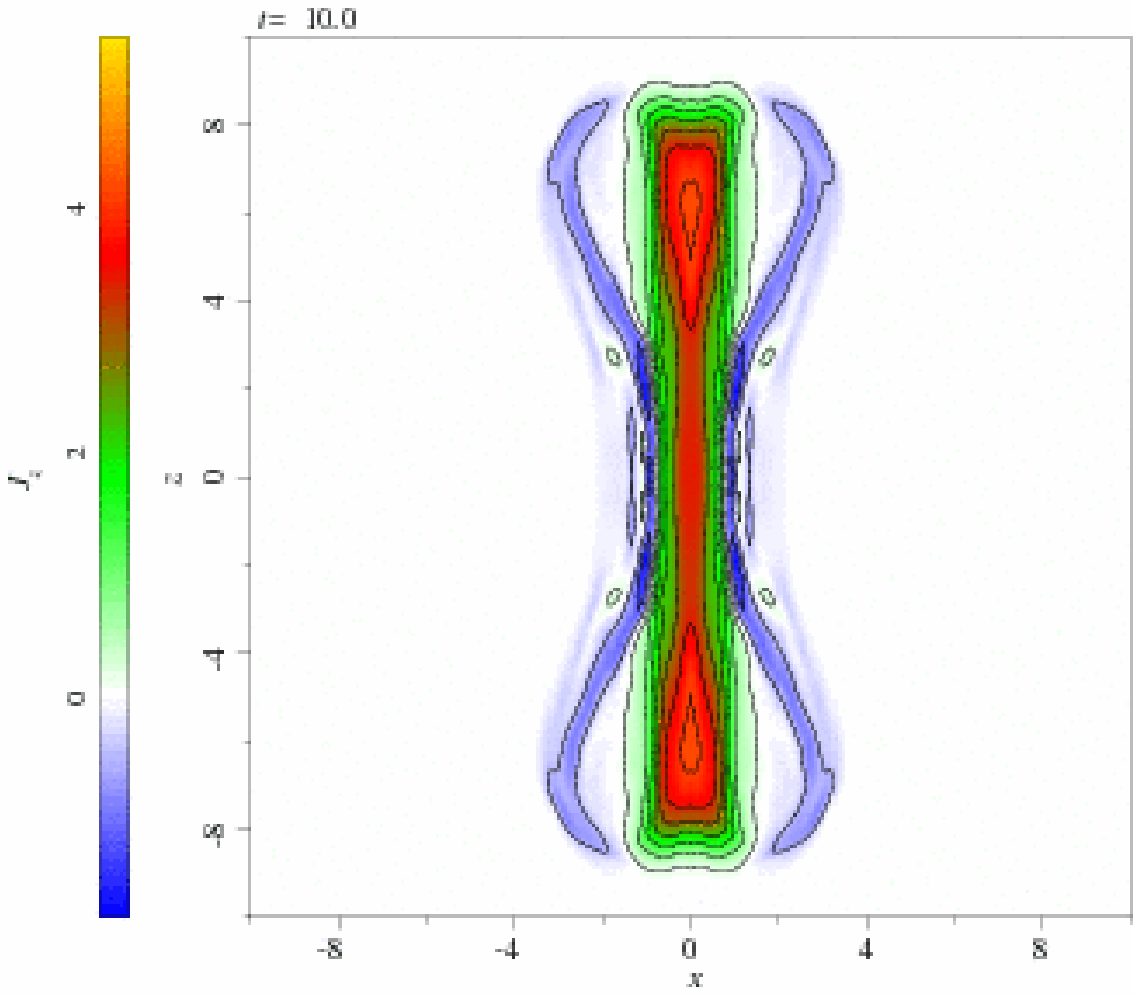}
\caption{\label{fig:final_Jz}   Distribution  of  the   axial  current
density  $J_z$ at  $t=10.0$  in the  $x-z$  plane. It  shows a  closed
circulating  current system,  in  which the  current  flows along  the
central  ($z$) column  ($J_z >  0$)  and returns  along the  conically
shaped shell on the outside ($J_z < 0$).}
\end{center}
\end{figure}

The   magnetic  tower   jet  has   a  well-ordered   helical  magnetic
configuration.  The 3-D view of  the selected magnetic lines of force,
as  illustrated in  Fig. \ref{fig:3Dlines},  indicates that  a tightly
wound central helix goes up along the central axis and a loosely wound
helix comes  back at the  outer edge of  the magnetic tower  jet.  The
magnetic pitch $B_{\phi}/\sqrt{B_r^2+B_z^2}$  has a broad distribution
with  a  maximum  of  $\sim  15$. Figure  \ref{fig:final_Jz}  shows  a
snapshot of the axial current density $J_z$ at $t=10.0$.  Clearly, the
axial  current flow displays  a closed  circulating current  system in
which  it  flows  along  the  central axis  (the  ``forward''  current
$\bolJ^{\rm F}$) and  returns on the conically shaped  path that is on
the  outside (the  ``return'' current  $\bolJ^{\rm R}$).   It  is well
known that an axial  current-carrying cylindrical plasma column with a
helical  magnetic field  is subject  to  current-driven instabilities,
such as sausage ($m=0$), kink  ($m=1$), and the other higher order $m$
modes ($m$ is  the azimuthal mode number).  We however  do not see any
visible evolution  of the  non-axisymmetric features in  this magnetic
tower jet.

From this overview,  we see that the magnetic  tower jet can propagate
through  the stratified background  medium while  keeping well-ordered
structures throughout  the time  evolution.  The magnetic  fields push
away the  background gas, forming  magnetically dominated, low-density
cavities. This  action also drives a  HD shock wave which  is ahead of
and eventually  separated from  the magnetic structures.  The magnetic
tower has  a slender ''body''  from the confinement of  the background
pressure  and an  expanded  ``lobe''  when the  fields  expand into  a
background with the decreasing pressure.

We will now turn to the  discussions on the detailed properties of the
tower jet,  including the HD  shock wave and  its impact in  the axial
($z$)   and   radial   ($x$)   directions  in   \S   \ref{sec:A}   and
\ref{sec:B}. The  radial force  balance of the  jet is examined  in \S
\ref{sec:C}.

\subsection{Structure of a Magnetic Tower Jet in the Axial ($z$) Direction}
\label{sec:A}

\begin{figure}
\begin{center}
\includegraphics[scale=0.9, angle=0]{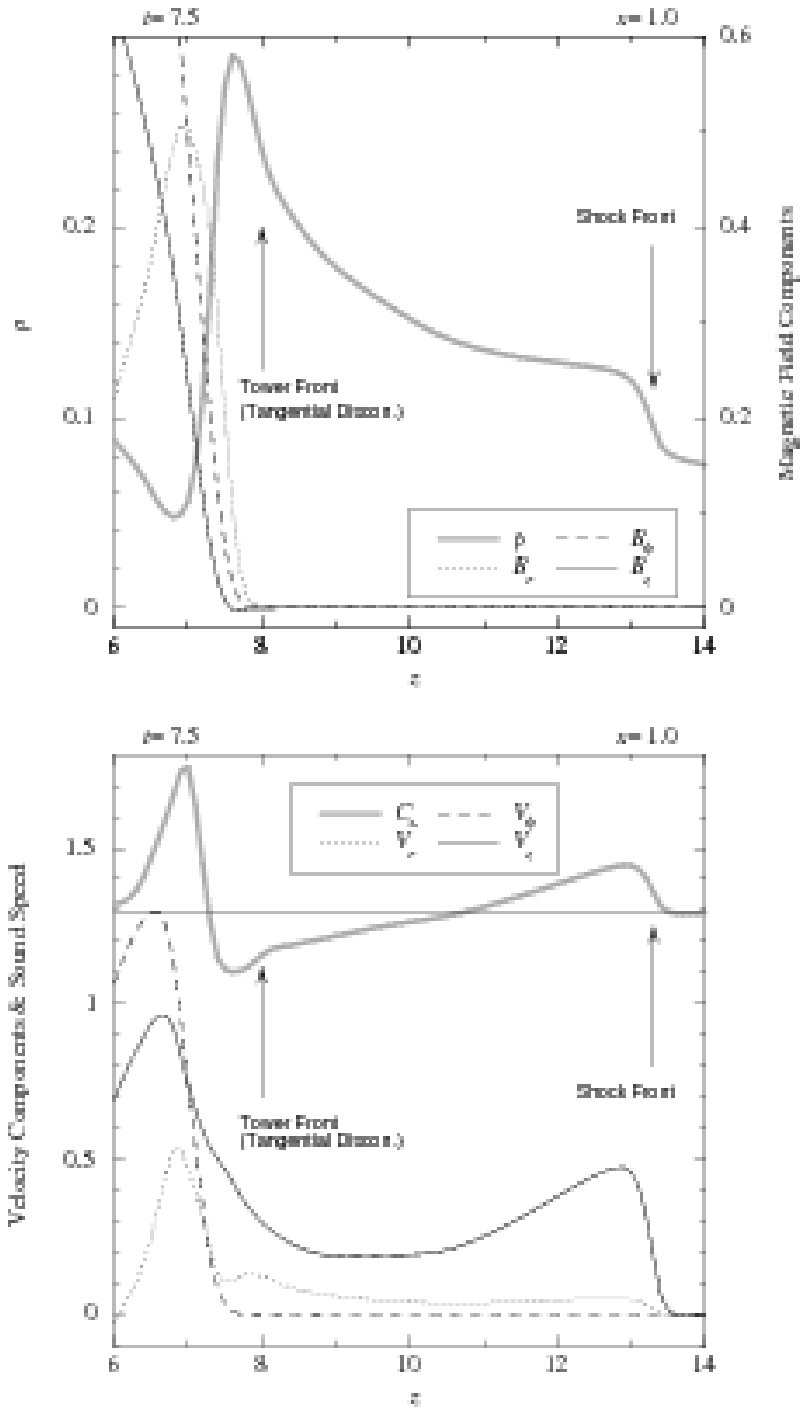}
\caption{\label{fig:t7.5_linez_1}    Axial   profiles    of   physical
quantities,  parallel  to   the  $z$-axis  with  $(x,\,y)=(1,\,0)$  at
$t=7.5$.   {\em Top}:  Density  $\rho$ and  magnetic field  components
$(B_r,\,B_\phi,\,B_z)$.  {\em  Bottom}: Sound  speed  $C_{\rm s}$  and
velocity  components  $(V_r,\,V_\phi,\,V_z)$.   The positions  of  the
expanding hydrodynamic shock wave  front and the magnetic tower front,
which is identified  as a tangential discontinuity, are  shown in both
panels.   A horizontal  {\em solid  line}  in the  {\em bottom}  panel
denotes the initial sound speed (constant throughout the computational
domain).  }
\end{center}
\end{figure}

Figure  \ref{fig:t7.5_linez_1}  displays  several physical  quantities
along a  line with $(x,y)=(1,0)$  in the axial direction  at $t=0.75$.
Several features can be identified. First, the HD shock wave front can
be seen around $z \sim 13.5$ in the profiles of $\rho$ ({\em top}
panel), $V_{z}$, and $C_{\rm s}$ ({\em bottom} panel).  $C_{\rm s}$ is
higher than the initial background value $1.29$ in the post-shock
region due to shock heating and becomes smaller than $1.29$ at $z \sim
10.7$ due to axial expansion.  $V_{z}$ has the similar behavior as
$C_{\rm s}$.

Second,  a magnetic  tower  front (``tower  front''  in the  following
discussions) is  located at  $z \sim 8.0$,  beyond which  the magnetic
field goes  to zero, as seen  in the {\em top}  panel.  This indicates
that  the gas  within the  magnetic tower  jet is  separated  from the
non-magnetized  ambient gas beyond  the tower  front.  We  regard this
front  as  a  tangential  discontinuity  as the  magnetic  fields  are
tangential  to  the  front  without  the  normal  component.  This  is
consistent  with  the  fact   that  the  radial  and  azimuthal  field
components ($B_{r}$ and $B_{\phi}$)  are dominant near the tower front
($z  \lesssim 8.0$)  but  the axial  field  component $B_{z}$  becomes
dominant  only  for $z<6.0$.  The  density  and  pressure show  smooth
transition  through this  front  though the  gradients  of $\rho$  and
$C_{\rm s}$ are slightly changed there.

\begin{figure}
\begin{center}
\includegraphics[scale=0.8, angle=0]{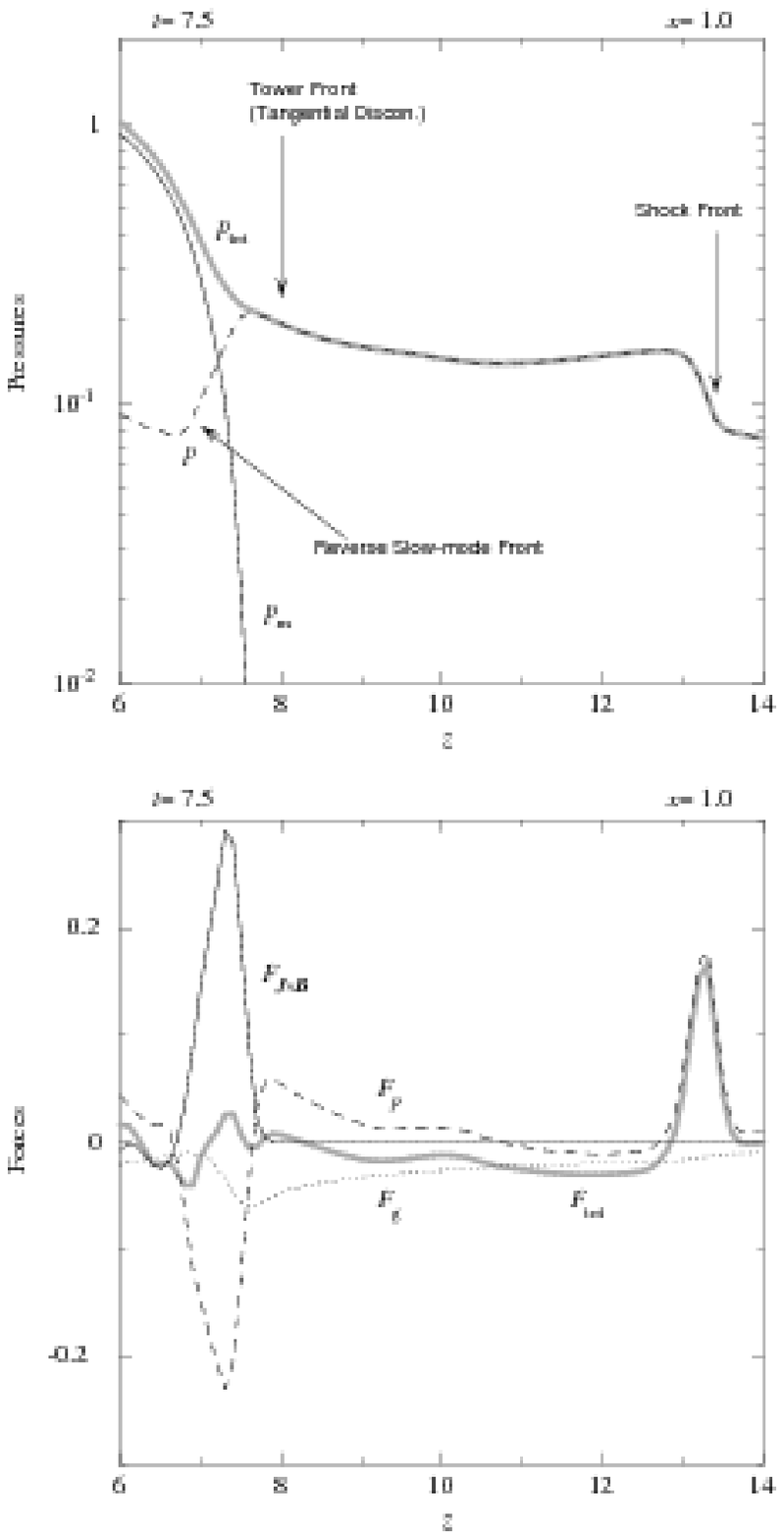}
\caption{\label{fig:t7.5_linez_2}              Similar              to
Fig. \ref{fig:t7.5_linez_1}.   {\em Top}:  Shown are the  gas pressure
$p$ ({\em dashed line}),  magnetic pressure ({\em solid line}) $p_{\rm
m}$, and total pressure $p_{\rm tot}\,(=p+p_{\rm m})$ ({\em light gray
thick solid line}).   {\em Bottom}: Shown are the  forces in the axial
($z$)    direction:    the     Lorentz    force    $F_{\bolJ    \times
\bolB}=-\partial/\partial z \left[(B_r^2+B_{\phi}^2)/2 \right]$, ({\em
solid line}), gas pressure  gradient force $F_p = -\partial p/\partial
z$ ({\em dashed line}),  gravitational force $F_{\rm g}=-\rho \partial
\psi / \partial  R \times |z|/R$ ({\em dotted  line}), and total force
$F_{\rm  tot}\,(=F_{\bolJ \times  \bolB}+F_p+F_{\rm  g})$ ({\em  light
gray thick  solid line}).  The  position of the reverse  slow-mode MHD
wave front is also shown in the {\em top} panel.  }
\end{center}
\end{figure}

\begin{figure}
\begin{center}
\includegraphics[scale=0.8, angle=0]{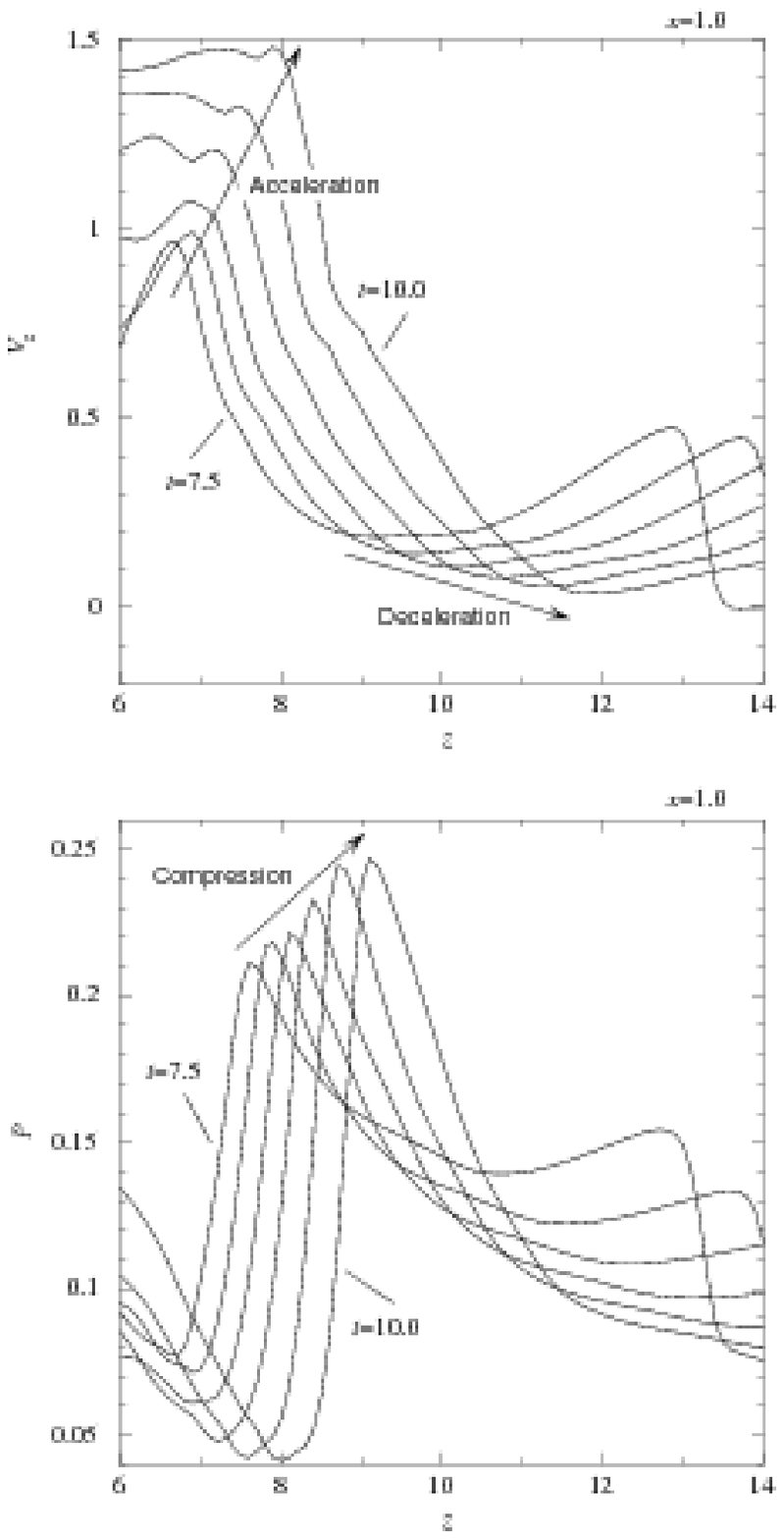}
\caption{\label{fig:t7.5_linez_3}              Similar              to
Fig. \ref{fig:t7.5_linez_1}, but with selected snapshots during $t=7.5
\sim 10.0$ (each time-interval is equal to 0.5).  {\em Top}: The axial
velocity $V_z$. {\em Bottom}: The gas pressure $p$.  }
\end{center}
\end{figure}

Third, there  is another MHD wave  front at $z \sim  7.0$ where $B_{r}$,
$C_{\rm  s}$,  and every  velocity  component  have  their local  maxima
($\rho$  instead has its  local minimum),  as seen  in both  panels.  To
better understand the  nature of this MHD wave front,  we plot the axial
profiles of pressures and various  forces along the line $(x,y) = (1,0)$
at $t=7.5$  in Fig. \ref{fig:t7.5_linez_2}.  The  total pressure $p_{\rm
tot}$ consists of  only the gas pressure $p$ beyond  the tower front ($z
\gtrsim  8.0$) but  is dominated  by the  magnetic pressure  $p_{\rm m}$
behind the MHD  wave front ($z \lesssim 7.0$), as seen  in the {\em top}
panel.  A transition occurs around  $7.0 \lesssim z \lesssim 8.0$, where
an increase  in $p$  is accompanied  by a decrease  in $p_{\rm  m}$.  We
therefore identify  this as a  reverse slow-mode compressional  MHD wave
front.  In  magnetic  towers,  the  transition region  between  gas  and
magnetic pressures can be identified as a reverse slow wave front in the
context of  MHD wave structures.  It  does not depend  on the resolution
and parameters.   So, the reverse slow  mode wave (sometimes,  it can be
steepen   into  a   shock)  will   always  be   there. In   addition, in
Fig. \ref{fig:t7.5_linez_3}, we show  several snapshots of $V_z$ and $p$
during $t=7.5  \sim 10.0$ (along the  same offset axial  path with Figs.
\ref{fig:t7.5_linez_1} and  \ref{fig:t7.5_linez_2}).  The axial  flow is
decelerated by  the gravitational force in the  post-shock region beyond
the tower front as seen in the  {\em top} panel.  On the other hand, the
narrow region  between the  tower front and  the reverse  slow-mode wave
front is  accelerated by the magnetic pressure  gradient (the ``magnetic
piston''  effect).  Note  that the  reverse slow-mode  MHD compressional
wave could eventually steepen into  the reverse slow-mode MHD shock wave
via this nonlinear evolution.  Consequently, in the frame co-moving with
the  reverse slow-mode  shock, a  strong compression  occurs  behind the
shock wave front and causes a local heating, as seen in the {\em bottom}
panel  and also  Fig. \ref{fig:final_prdiff}.   This heating  could have
interesting implications for the  enhancement of radiation from radio to
X-rays at the terminal part of  Fanaroff-Riley type II AGN jets, such as
lobes and hot  spots, which are generally interpreted  as heating caused
by the jet terminal shock wave \citep[]{BR74, S74}.

Fourth, the HD shock wave breaks the initial background hydrostatic
equilibrium. The passage of the shock wave heats the gas and alters
its pressure gradient.  As shown in the {\em bottom} panel of
Fig. \ref{fig:t7.5_linez_2}, the gas pressure gradient force $F_{\rm
p}$ stays uupositive at the shock front (which pushes the shock
forward), but the total (gravity plus pressure gradient) force $F_{\rm
tot}$ becomes negative behind the shock, implying a deceleration of
the gas in the axial direction in the post-shock region. This is
consistent with Fig. \ref{fig:t7.5_linez_3}.

\subsection{Deformation of the Jet ``Body'' in the Radial Direction}
\label{sec:B}

\begin{figure}
\begin{center}
\includegraphics[scale=0.9, angle=0]{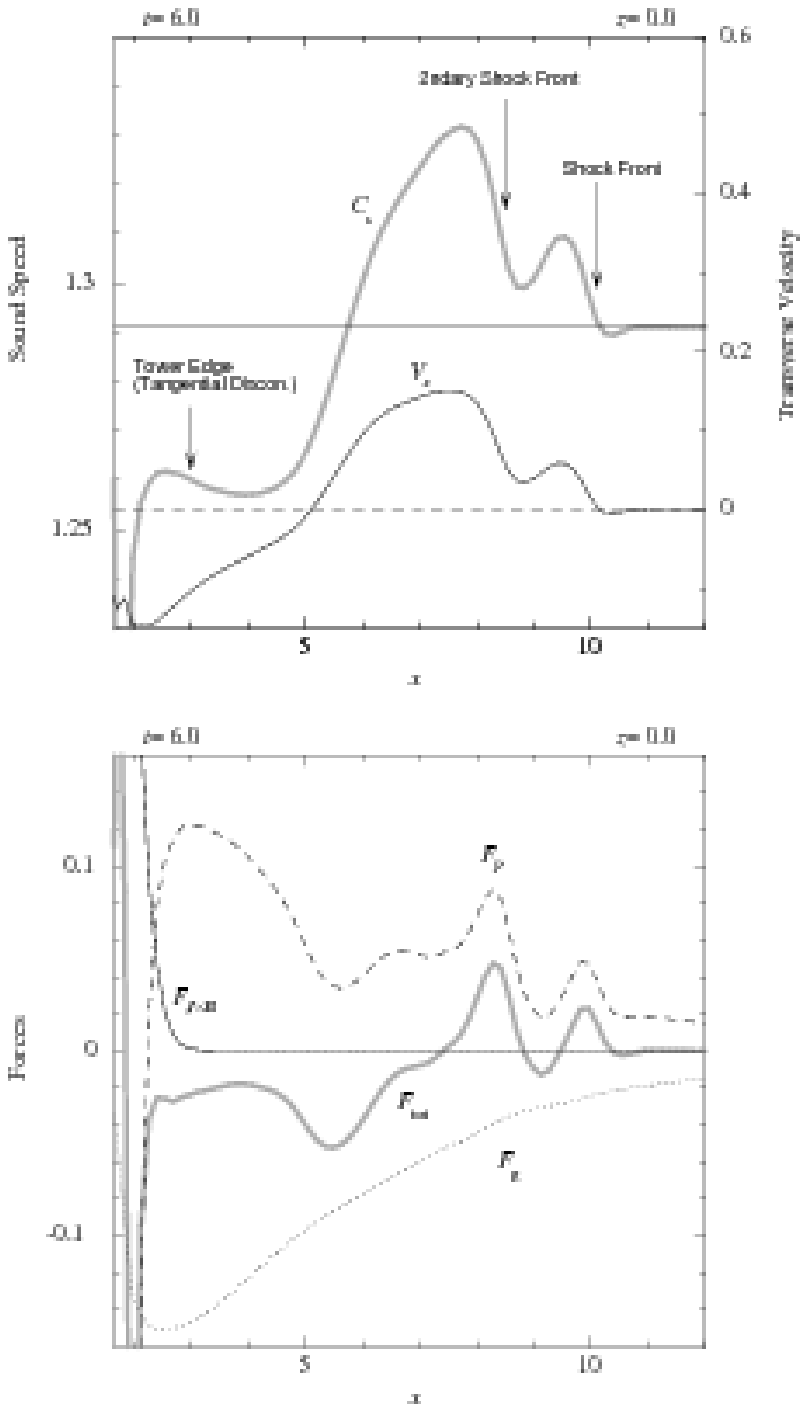}
\caption{\label{fig:t6.0_liner} Radial profiles of physical quantities
along the  $x$-axis in the equatorial plane  with $(y,\,z)=(0,\,0)$ at
$t=6.0$.   {\em Top}:  The sound  speed $C_{\rm  s}$ ({\em  light gray
thick solid line})  and the radial velocity $V_x$  ({\em solid line}).
{\em Bottom}:  The forces in  the radial ($x$) direction:  the Lorentz
force      $F_{\bolJ      \times      \bolB}=-\partial/\partial      r
\left[(B_{\phi}^2+B_{z}^2)/2\right]-B_{\phi}^2/r$  ({\em solid line}),
gas pressure  gradient force  $F_p = -\partial  p / \partial  r$ ({\em
dashed line}),  gravitational force  $F_{\rm g}=-\rho \partial  \psi /
\partial R \times |x|/R$ ({\em  dotted line}), and total force $F_{\rm
tot}\,(=F_{\bolJ \times \bolB}+F_p+F_{\rm  g})$ ({\em light gray thick
solid line}).  The positions  of two expanding hydrodynamic shock wave
fronts and  the edge of the magnetic  tower (tangential discontinuity)
are shown  in the {\em top}  panel.  A horizontal {\em  solid line} in
the  {\em  top}  panel  denotes  the  initial  sound  speed  (constant
throughout  the computational  domain)  and a  horizontal {\em  dashed
line}  in  the  {\em top}  panel  represents  a  level ``0''  for  the
transverse velocity.  }
\end{center}
\end{figure}

We next examine  the structure and dynamics of  the magnetic tower jet
along the radial direction in the equatorial plane with $(y,z)=(0,0)$.
Figure  \ref{fig:t6.0_liner}  shows the  radial  profiles of  physical
quantities  along  the  $x$-axis  at  $t=6.0$.  The  boundary  of  the
magnetic tower  jet (``tower edge''  in the following  discussions) is
located at $x  \sim 3.0$ where $F_{\bolJ \times  \bolB}$ becomes zero.
Two distinct peaks of $C_{\rm s}$  and $V_{x}$ around $x \sim$ 7.8 and
9.5 are  visible in  the {\em  top} panel.  The  first front  ($x \sim
9.5$)  is the  propagating HD  shock wave  front as  we showed  in the
previous  section.  The  second front  ($x \sim  7.8$)  also indicates
another expanding HD  shock wave front generated by  a bounce when the
magnetic flux  pinches in  the radial direction  caused by  the ``hoop
stress''.   This secondary  shock  front appears  only  in the  radial
direction  (see   also  Fig.  \ref{fig:t7.5_linez_1}).    $C_{\rm  s}$
decreases gradually towards  the jet axis in the  post-shock region of
the secondary shock and becomes  smaller than its initial value $1.29$
at $x  \sim 5.7$.  We can  confirm that these shock  fronts are purely
powered by  the gas  pressure gradients $F_{p}$,  as seen in  the {\em
bottom} panel (twin peaks of $F_{\rm p}$ are shifted a bit behind that
of $C_{\rm s}$ in the {\em top} panel).

Part of the region between the  secondary HD shock front and the tower
edge  has  $F_{\rm tot}$  being  negative  (the  {\rm bottom}  panel),
meaning that the gas  is undergoing gravitational contraction. This is
indicated by  $V_x < 0$ in  the {\em top}  panel for $x <  5.1$.  This
behavior is  similar to what we  have discussed earlier,  i.e., the HD
shock waves  break the  background hydrostatic equilibrium,  causing a
global contraction. Note that  this contraction is occurring along the
whole jet ``body'', e.g., for $|z| \lesssim 4.0$ when $t=7.5$.

The  {\em bottom} panel  also helps  to address  the question  on what
forces  are confining  the magnetic  fields in  the  equatorial plane.
Since  the  total  magnetic   field  $F_{\bolJ  \times  \bolB}$  stays
positive, this means that the  inward hoop stress is not strong enough
to confine the magnetic fields. Instead,  at the tower edge, it is the
background  gravity that  counters  the combined  effects of  magnetic
field pressure  and a positive pressure gradient  (pushing outward). A
bit  further into  the  tower edge  (at  $x \sim  2.1$), however,  the
pressure   force   changes  from   outwardly   directed  to   inwardly
directed.  Then,  both the  pressure  gradient  and the  gravitational
forces act to  counter the outward ${\bolJ \times  \bolB}$ force. This
behavior, which is mostly caused  by the magnetic tower expanding in a
background gravitational field, is different from the usual MHD models
for jets  where the inwardly directed  hoop stress is  balanced by the
outwardly directed magnetic pressure gradient \citep[]{BP82}.

\begin{figure}
\begin{center}
\includegraphics[scale=0.4, angle=0]{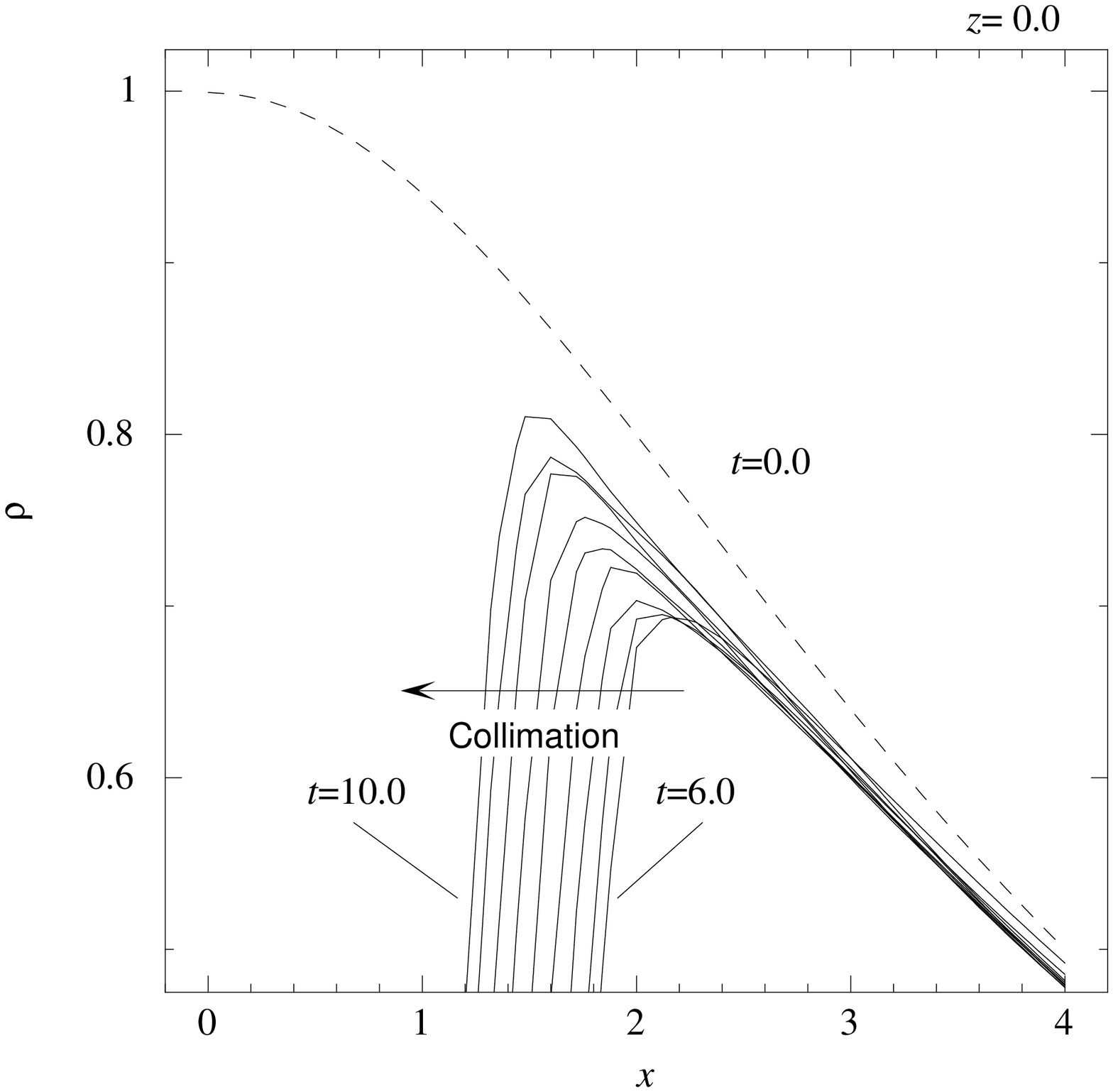}
\caption{\label{fig:liner_de} The radial profiles of density $\rho$ in
the equatorial plane with  selected snapshots during $t=6.0 \sim 10.0$
(each time-interval  is equal to 0.5).  The  initial profile ($t=0.0$)
is also shown ({\em dashed line}).  }
\end{center}
\end{figure}

To investigate the radius evolution of the magnetic tower jet, we show
the radial profile of density  at the equatorial plane from $t=6.0$ to
$ 10.0$ in  Fig. \ref{fig:liner_de}.  It shows that  the radius has an
approximately constant  contraction speed  for $t=6.5 \sim  9.5$.  The
time  scale for  contraction is  $\tau_{\rm contr}  \sim 6$,  which is
about 7.5 times longer than the local sound-crossing time scale.

\subsection{Force Balance in the Radial Direction}
\label{sec:C}

\begin{figure}
\begin{center}
\includegraphics[scale=0.8, angle=0]{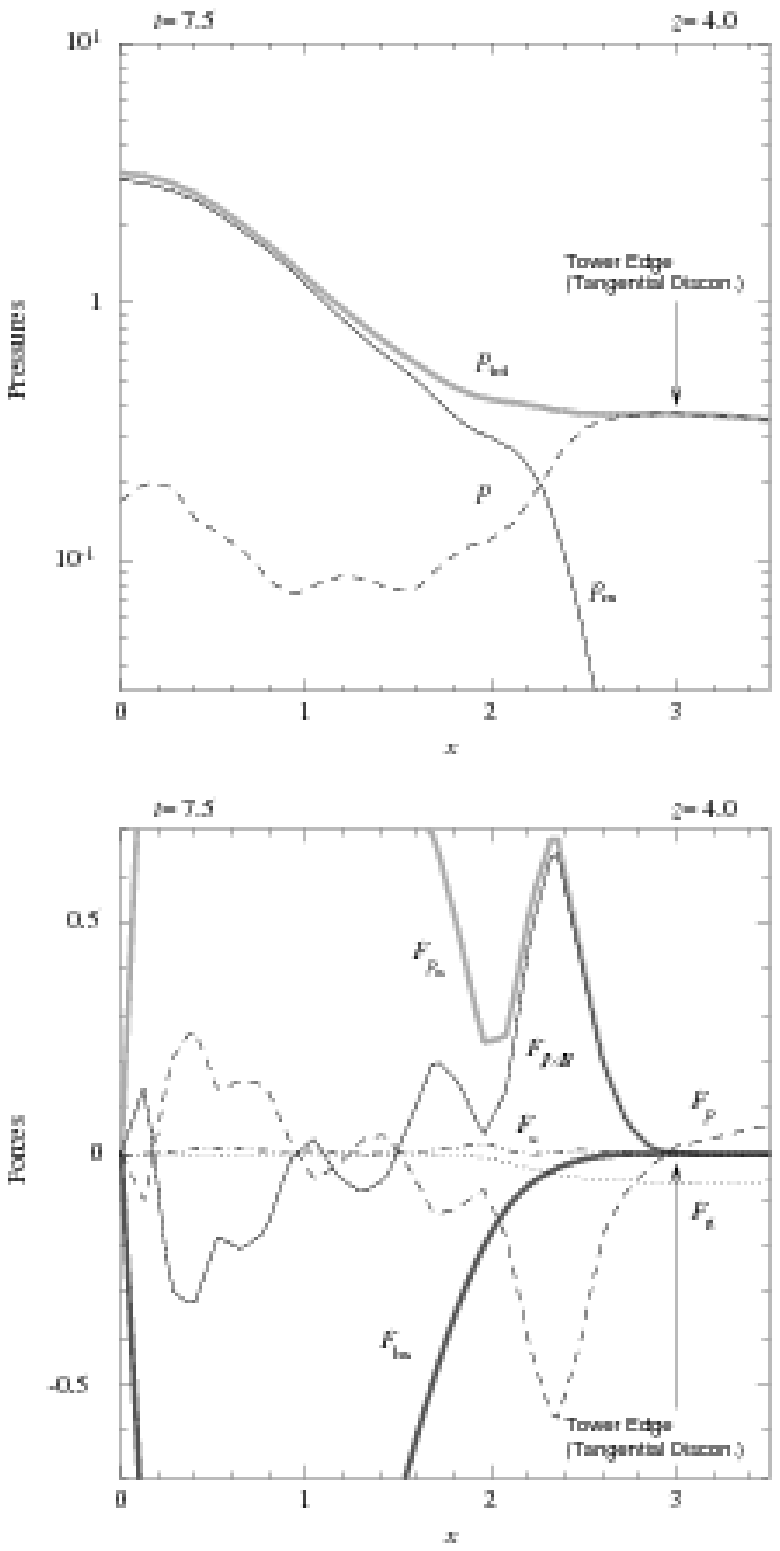}
\caption{\label{fig:t7.5_liner}   The  radial  profiles   of  physical
quantities along the $x$-axis with $(y,\,z)=(0,\,4)$ at $t=7.5$.  {\em
Top}: Similar  to the {\em top} panel  in Fig. \ref{fig:t7.5_linez_2}.
{\em    Bottom}:   Similar    to   the    {\em   bottom}    panel   in
Fig.  \ref{fig:t6.0_liner},  but with  the  centrifugal force  $F_{\rm
c}=\rho V_{\phi}^2/r$  ({\em dash-dotted line}) and  the Lorentz force
$F_{\bolJ  \times \bolB}$, which  can be  separated into  the magnetic
pressure       gradient        force       $-\partial/\partial       r
\left[(B_{\phi}^2+B_{z}^2)/2\right]$  ({\em  light  gray  thick  solid
line}) and  the hoop  stress (magnetic tension  force) $-B_{\phi}^2/r$
({\em  dark gray  thick solid  line}).  The  position of  the magnetic
tower edge (tangential discontinuity) is shown in both panels.  }
\end{center}
\end{figure}

We now discuss the jet properties along the radial direction away from
the  equatorial plane.  Figure  \ref{fig:t7.5_liner} shows  the radial
profiles of physical quantities  along the $x$-axis with $(y,z)=(0,4)$
at  $t=7.5$.  The  tower edge  is now  located at  $x \sim  3.0$.  The
plasma $\beta$ in the core of  the tower is $\beta \lesssim 0.1$.  The
{\em  top} panel  shows  that  the central  total  pressure (which  is
dominated by the magnetic pressure) is much bigger than the background
``confining'' pressure.   This illustrates the  original suggestion by
Lynden-Bell \citep[]{L96,  L03} that the  hoop stress of  the toroidal
field  component $B_{\phi}$  can act  as a  pressure amplifier  in the
central  region of  the  magnetic tower:  the  pinch effect  amplifies
$p_{m}$ near  the axis  of the  tower. At the  tower edge,  however, a
finite   (albeit   small)   gas   pressure  is   required   \citep[see
also][]{Li01}.

\begin{figure}
\begin{center}
\includegraphics[scale=0.4, angle=0]{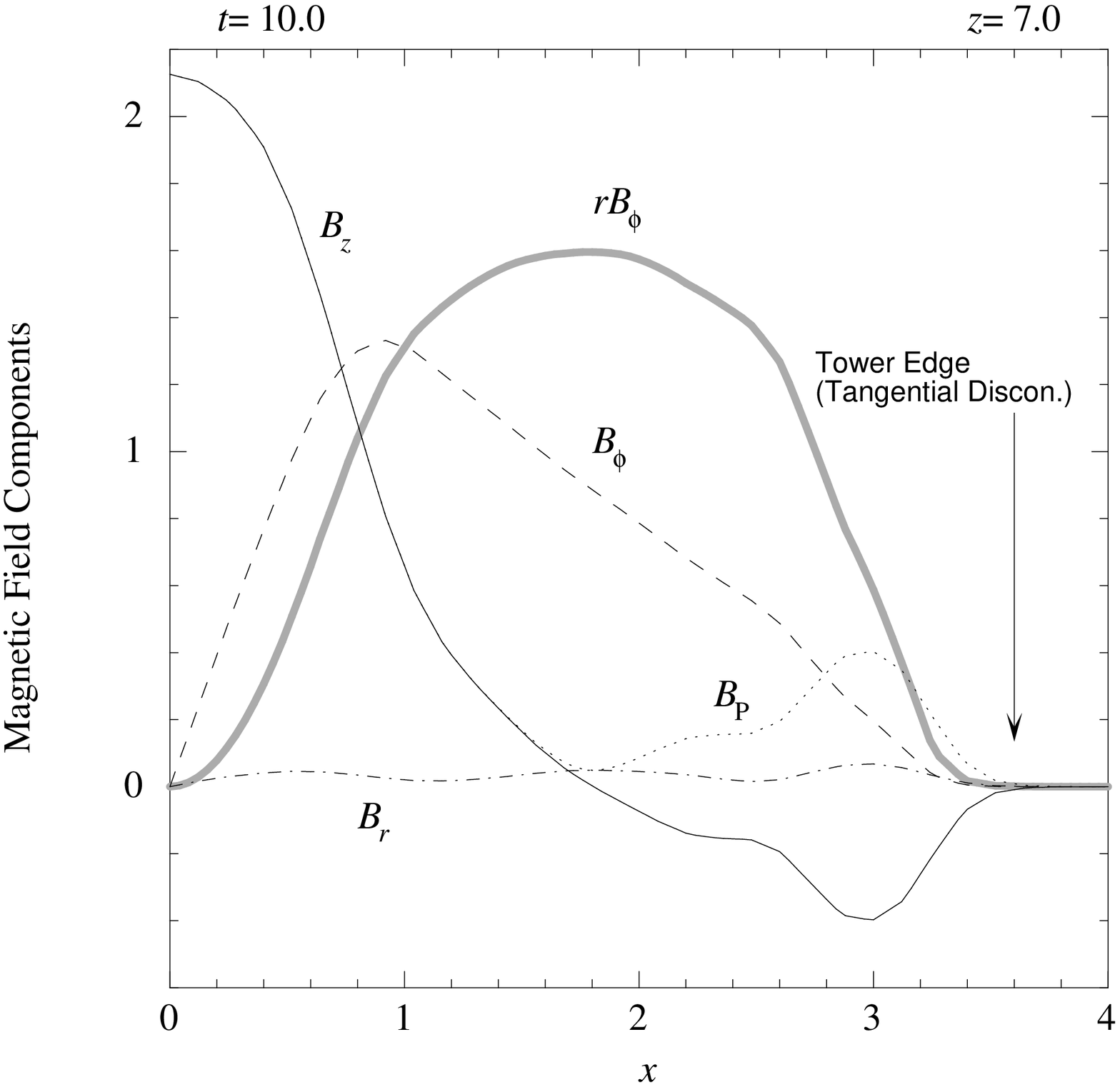}
\caption{\label{fig:t10.0_liner_B} The radial profiles of the magnetic
field components  $(B_r,\,B_\phi,\,B_z)$, the poloidal  magnetic field
$B_{\rm  p}=\sqrt{B_r^2+B_{z}^2}$, and the  quantity $r  B_\phi$ along
the $x$-axis  with $(y,\,z)=(0,\,7)$ at $t=10.0$. The  position of the
magnetic tower edge (tangential discontinuity) is shown.  }
\end{center}
\end{figure}

The  {\em bottom}  panel shows  the detailed  distributions  of forces
along the radial direction. They show that:
\begin{itemize}
\item{(Region: $x \gtrsim 3.0$) Beyond the tower edge, the
gravitational force $F_{\rm g}$ is slightly stronger than the
outwardly directed gas pressure gradient force $F_{p}$, indicating
that this edge is subject to the gravitational contraction, as
discussed in the previous section.}
\item{(Region: $2.0 \lesssim x \lesssim 3.0$) Interior to the tower
edge, the Lorentz force $F_{\bolJ \times \bolB}$} is dominated by the
outwardly directed magnetic pressure gradient force $F_{p_{m}}$, and
it is also larger than the inwardly directed $F_{p}$, indicating that
the outer shell of the magnetic shell should be expanding at the
relatively higher $z$, in contrast to the equatorial region ($z=0$)
where the tower radius is contracting.  The hoop stress $F_{\rm hp}$
plays a minor role in the force balance around this region.  Note that
$\bolJ^{\rm R}$ flows inside this region.
\item{(Region: $x \lesssim 2.0$) Inside the jet ``body'', 
Contributions from both $F_{p_{m}}$ and $F_{\rm hp}$ 
become comparable and nearly cancel each other. 
The residual $F_{\bolJ \times \bolB}$ is balanced by
$F_{p}$ everywhere in this region. 
$F_{\rm hp}$ becomes dominant in the Lorentz force at the core part.
Note that $\bolJ^{\rm F}$ flows within $x \lesssim 1.0$.}
\end{itemize} 
In addition,  both the $F_{\rm  g}$ and the centrifugal  force $F_{\rm
c}$  play a  minor  role in  terms  of the  force  balance inside  the
magnetic  tower jet.   Thus, the  interior  of the  magnetic tower  is
magnetically dominated but not exactly force-free, i.e., $- \nabla p +
\bolJ  \times  \bolB \simeq  0~~$.   This  small  but finite  pressure
gradient force  could potentially provide some  stabilizing effects on
the traditionally kink-unstable  twisted magnetic configurations.  The
detailed examinations  of stability properties in  magnetic tower jets
will be discussed in a forthcoming paper.

\section{DISCUSSIONS}

On scales  of $\sim$ tens  of kpc to  hundreds of kpc,  the background
density and pressure  profiles have a strong influence  on the overall
morphology  of the magnetic  tower jet.  Most notably,  the transverse
size  of  the  magnetic tower  grows  as  the  jet propagates  into  a
decreasing pressure environment, showing a similar morphology with the
jet/lobe configuration of radio galaxies.

The radial size  of the lobe $r_{\rm lobe}$ can  be estimated from the
following  consideration:  Figure  \ref{fig:t10.0_liner_B}  shows  the
radial profiles of magnetic  field components parallel to the $x$-axis
with $z=7.0$  at $t=10.0$.  Together with Fig.  \ref{fig:final_Jz}, we
see that both  the poloidal field and especially  the poloidal current
$I_z$ maintain  well collimated  around the central  axis even  at the
late stage of  the evolution. This implies that  the toroidal magnetic
fields  in the  lobe region  are distributed  roughly as  $B_\phi \sim
I_z/r$.    This   is   consistent    with   the   results   shown   in
Fig.  \ref{fig:t10.0_liner_B}  where  $rB_\phi$  have a  plateau  from
$x\approx 1-2.5$.  As indicated  in Fig.  \ref{fig:t7.5_liner}, we see
that the magnetic pressure and  the background pressure try to balance
each other at the tower edge. Thus, we expect that
\begin{equation}
\label{eq:balance_edge}
\frac{B_{\rm P}^{2}+B_{\phi}^{2}}{8 \pi} \sim p_{\rm e}~,
\end{equation} 
where $p_{\rm e}$ is the external gas pressure at the tower edge. 
When the lobe experiences sufficient expansion, we expect the poloidal
field strength to drop much faster than $1/r_{\rm lobe}$. So we have
\begin{equation}
\label{eq:balance_edge2}
\frac{B_{\phi}^{2}}{8 \pi} \sim \frac{(I_z/r_{\rm
lobe})^2}{8\pi}\sim p_{\rm e}~, 
\end{equation} 
which gives 
\begin{equation}
r_{\rm lobe} \sim I_z~p_e^{-1/2}~~.  
\end{equation} 
This is generally  consistent with our numerical result  shown in Fig.
\ref{fig:de_evo} though  it is difficult  to make a  firm quantitative
determination since the lobes have not expanded sufficiently.

To make  direct comparison  between our simulations  and observations,
further analysis is  clearly needed. Note that for  the magnetic tower
model,  the Alfv\'en  surface  is located  at  the outer  edge of  the
magnetic  tower   and  flow  within  the  magnetic   tower  is  always
sub-Alfv\'enic. This is quite  different from the hydromagnetic models
where  the MHD  flow is  accelerated  and has  passed through  several
critical velocity surfaces, including the Alfv\'en surface.

\section{CONCLUSION}

By performing 3-D  MHD simulations we have investigated  in detail the
nonlinear dynamics of magnetic tower jets, which propagate through the
stably  stratified   background  that  is   initially  in  hydrostatic
equilibrium.   Our simulations,  based  on the  approach developed  in
\citet{Li06}, confirm a number of the global characteristics developed
in  Lynden-Bell  (1996, 2003)  and  Li  et  al. (2001).   The  results
presented here give additional  details for a dynamically evolving jet
in a  stratified background.   The magnetic tower  is made  of helical
magnetic fields, with poloidal  flux and poloidal current concentrated
around  the central axis.  The ``returning''  portion of  the poloidal
flux and  current is  distributed on the  outer shell of  the magnetic
tower. Together they form a self-contained system with magnetic fields
and associated  currents, being confined  by the ambient  pressure and
gravity.  The  overall morphology  exhibits  a  confined magnetic  jet
``body'' with  an expanded ``lobe''.  The confinement of  the ``body''
comes  jointly  from  the  external  pressure  and  the  gravity.  The
formation of the lobe is due  to the expansion of magnetic fields in a
decreasing background pressure.

A  hydrodynamic shock  wave  initiated by  the  injection of  magnetic
energy/flux propagates ahead  of the magnetic tower and  can break the
hydrostatic  equilibrium  of  the  ambient medium,  causing  a  global
gravitational contraction.   As a result, a  strong compression occurs
in  the axial  direction  between  the magnetic  tower  front and  the
reverse slow-mode MHD shock wave front that follows.  Furthermore, the
magnetic tower  jets are deformed radially into  a slender-shaped body
due to the inward-directed flow of the ambient (non-magnetized) gas.

The  lobe is  magnetically dominated  and  is likely  filled with  the
toroidal magnetic  fields generated  by the central  poloidal current.
At the edge  of the magnetic tower jet,  the outward-directed magnetic
pressure  gradient  force  is  balanced  by  the  inward-directed  gas
pressure gradient force, so the radial width of the magnetic tower can
be determined jointly by the magnitude of the poloidal current and the
magnitude  of the  external gas  pressure.  The  highly  wound helical
magnetic  field in  the magnetic  tower never  reaches  the force-free
equilibrium precisely, but obtains radial force-balance, including the
gas pressure gradient inside the magnetic tower.

The stability  of the  magnetic tower jets  will be considered  in our
forthcoming papers.

\acknowledgments 

Useful discussions with John Finn, Stirling Colgate and Ken Fowler are
gratefully acknowledged.  This work was carried out under the auspices
of the National Nuclear Security Administration of the U.S. Department
of   Energy  at   Los  Alamos   National  Laboratory   under  Contract
No. DE-AC52-06NA25396.   It was  supported by the  Laboratory Directed
Research and Development Program at LANL and by IGPP at LANL.

\begin{deluxetable}{rlll}
\tabletypesize{\scriptsize}
\tablecaption{Units of Physical Quantities for Normalization. \label{tbl:unit}}
\tablewidth{0pt}
\tablehead{
\colhead{Physical Quantities} & \colhead{Description}   &
 \colhead{Normalization Units} & \colhead{Typical Values}}
\startdata
$R\,(= \sqrt{x^2+y^2+z^2})$\dotfill & Length & $R_{0}  $ & 5 Kpc  \\ 
${\bolV}$\dotfill & Velocity field & $C_{\rm s0}            $ & $4.6 \times 10^7$ cm/s\\
$t$\dotfill       & Time      & $R_{\rm 0}/C_{\rm s0}       $ & $1.0 \times 10^7$ yrs\\
$\rho$\dotfill    & Density   & $\rho_{0}                   $ & $5.0 \times 10^{-27}$ g/cm$^3$\\
$p$\dotfill       & Pressure  & $\rho_{0} C_{\rm s0}^{2}    $ & $1.4 \times 10^{-11}$ dyn/cm$^2$\\
${\bolB}$\dotfill & Magnetic field  & $\sqrt{4 \pi \rho_{0} C^{2}_{\rm s0}}  $ & 17.1 $\mu$G\\
\enddata
\tablecomments{
The initial  value of the  density $\rho_{0}$ and sound  speed $C_{\rm
s0}$  at the  origin $(x,\,y,\,z)=(0,\,0,\,0)$  are chosen  to  be the
typical density and velocity in the system.  The initial dimensionless
density $\rho^{\prime}$  and pressure  $p^{\prime}$ at the  origin are
set to unity.  We therefore have the initial dimensionless sound speed
$C_{\rm s}^{\prime}=\gamma^{1/2} \approx 1.29$.  A characteristic time
scale, the initial sound crossing  time $\tau_{\rm s 0} = R_{0}/C_{\rm
s  0}=10.0$ Myrs, which  corresponds to  the dimensionless  time scale
$\tau \approx 0.78$.  }
\end{deluxetable}


\begin{thebibliography}{}
\bibitem[Blandford \& Rees (1974)]{BR74}
        Blandford, R. D., \& Rees, M. J. 1974, \mnras, 169, 395
\bibitem[Blandford (1976)]{B76}
        Blandford, R. D. 1976, \mnras, 199, 883
\bibitem[Blandford \& Znajek (1977)]{BZ77}
        Blandford, R. D., \& Znajek, R. L. 1977, \mnras, 179, 433
\bibitem[Blandford \& Payne (1982)]{BP82}
        Blandford, R. D., \& Payne, D. G. 1982, \mnras, 199, 883
\bibitem[Churazov et al. (2003)]{C03}
        Churazov, E., Forman, W., Jones, C., \& B\"{o}hringer, H. 2003, \apj, 590, 225
\bibitem[Ferrari (1998)]{F98}
        Ferrari, A. 1998, \araa, 36, 539
\bibitem[Hsu \& Bellan (2002)]{HB02}
        Hsu, S. C., \& Bellan, P. M. 2002, \mnras, 334, 257
\bibitem[Kato, Hayashi, \& Matsumoto (2004)]{K04a}
        Kato, Y., Hayashi, M. R., \& Matsumoto, R. 2004, \apj, 600, 338
\bibitem[Kato, Mineshige, \& Shibata (2004)]{K04b}
        Kato, Y., Mineshige, S., \& Shibata, K. 2004, \apj, 605, 307
\bibitem[King (1962)]{K62}
        King, I. 1962, \aj, 67, 471
\bibitem[Kudoh, Matsumoto \& Shibata (2002)]{K02}
        Kudoh, T., Matsumoto, R., \& Shibata, K. 2002, \pasj, 54, 26
\bibitem[Lebedev et al. (2005)]{L05}
        Lebedev, S. V. et al. 2005, \mnras, 361, 97
\bibitem[Li et al. (2001)]{Li01}
        Li, H., Lovelace, R. V. E., Finn, J. M., \& Colgate,
        S. A. 2001,  \apj, 561, 915
\bibitem[Li et al. (2006)]{Li06}
        Li, H., Lapenta, G., Finn, J. M., Li, S., \& Colgate, S. A. 
        2006, \apj~in press (astro-ph/0604469)
\bibitem[Li \& Li (2003)]{LL03}
        Li, S., \& Li, H. 2003, Technical Report, LA-UR-03-8935, 
        Los Alamos National Laboratory
\bibitem[Lovelace (1976)]{L76}
        Lovelace, R. V. E. 1976, \nat, 262, 649
\bibitem[Lovelace et al. (2002)]{L02}
        Lovelace, R. V. E., Li, H., Koldoba, A. V., 
        Ustyugova, G. V., \& Romanova, M. M. 2002, \apj, 572, 445
\bibitem[Lovelace \& Romanova (2003)]{LR03}
        Lovelace, R. V. E., \& Romanova, M. M. 2003, \apjl, 596, L159  
\bibitem[Lynden-Bell (1996)]{L96}
        Lynden-Bell, D. 1996, \mnras, 279, 389
\bibitem[Lynden-Bell (2003)]{L03}
        Lynden-Bell, D. 2003, \mnras, 341, 1360
\bibitem[Lynden-Bell (2006)]{L06}
        Lynden-Bell, D. 2006, \mnras~in press (astro-ph/0604424)
\bibitem[Lynden-Bell \& Boily (1994)]{LB94}
        Lynden-Bell, D., \& Boily, C. 1994, \mnras, 267, 146      
\bibitem[Meier, Koide, \& Uchida (2001)]{M01}
        Meier, D. L., Koide, S., \& Uchida, Y. 2001, Science, 291, 84
\bibitem[Romanova et al. (1998)]{R98}
        Romanova, M. M., Ustyugova, G. V., Koldoba, A. V., 
        Chechetkin, V. M., \& Lovelace, R. V. E. 1998, \apj, 500, 703
\bibitem[Scheuer (1974)]{S74}
        Scheuer, P. A. G. 1974, \mnras, 166, 513
\bibitem[Turner, Bodenheimer, \& R\'{o}\.{z}yczka (1999)]{T99}
        Turner, N. J., Bodenheimer, P., \& R\'{o}\.{z}yczka M. 1999, \apj, 524, 12
\bibitem[Uchida \& Shibata (1985)]{US85}
        Uchida, Y., \& Shibata, K. 1985, \pasj, 37, 515
\bibitem[Ustyugova et al. (2000)]{U00}
        Ustyugova, G. V., Lovelace, R. V. E., Romanova, M. M., Li, H.,
        Colgate, S. A. (2000), \apjl, 541, L21
\bibitem[Uzdensky \& MacFadyen (2006)]{UM06}
        Uzdensky, D. A., \& MacFadyen, A. I. 2006, preprint (astro-ph/0602419)
\end{thebibliography}
\end{document}